\documentclass[12pt]{article}

\usepackage{amsmath,amssymb,amsbsy,amsfonts,latexsym,graphicx}
\usepackage{color,array}

\allowdisplaybreaks

\evensidemargin \oddsidemargin

 \def\p{\partial}
 \def\f {\frac}
  
\def\nn{\nonumber}

\begin{document}


\begin{titlepage}

\renewcommand{\thefootnote}{\fnsymbol{footnote}}


\begin{flushright}
KIAS-P14022
\end{flushright}

\vspace{15mm}
\baselineskip 9mm
\begin{center}
  {\Large \bf Renormalized Entanglement Entropy Flow \\
  in Mass-deformed ABJM Theory}
\end{center}

\baselineskip 6mm
\vspace{10mm}
\begin{center}
 Kyung Kiu Kim,$^1$ O-Kab Kwon,$^2$ Chanyong Park,$^2$
 and Hyeonjoon Shin$^3$
 \\[10mm]
  $^1${\sl Department of Physics and Photon Science,
   School of Physics and Chemistry, \\ GIST, Gwangju 500-712, Korea} 
  \\[3mm] 
  $^2${\sl Institute for the Early Universe, Ewha Womans University, 
  Seoul 120-750, Korea}
  \\[3mm] 
  $^3${\sl School of Physics, Korea Institute for Advanced Study, 
  Seoul 130-722, Korea}
  \\[10mm]
  {\tt kimkyungkiu@gmail.com,~okabkwon@ewha.ac.kr,
cyong21@ewha.ac.kr,~hyeonjoon@kias.re.kr} 
\end{center}

\thispagestyle{empty}

\vfill
\begin{center}
{\bf Abstract}
\end{center}
\noindent
We investigate a mass deformation effect on the 
renormalized entanglement entropy (REE)  near the UV fixed point 
in (2+1)-dimensional field theory. 
In the context of the gauge/gravity duality, we use the 
Lin-Lunin-Maldacena (LLM) geometries corresponding
to the vacua of the mass-deformed ABJM theory. 
We analytically compute the small mass effect for various droplet configurations 
and show in holographic point of view 
that the REE is monotonically decreasing, positive, and stationary 
at the UV fixed point. 
These properties of the REE in (2+1)-dimensions are consistent with   
the Zamolodchikov $c$-function proposed in (1+1)-dimensional conformal field theory.  
\\ [15mm] 

\vspace{5mm}
\end{titlepage}

\baselineskip 6.6mm
\renewcommand{\thefootnote}{\arabic{footnote}}
\setcounter{footnote}{0}

\section{Introduction}
%
One of the well-known features of the entanglement entropy (EE) 
in quantum field theories 
is the appearance of the area law describing short range correlation. 
This correlation causes the UV divergence in the continuum limit and can be 
regulated  in terms of the UV cutoff~\cite{Bombelli:1986rw,Srednicki:1993im}. 
In other words, the EE is  UV sensitive. 
Due to this fact, it is not a good observable 
to measure the number of  degrees of freedom related to the long range correlations 
of the ground state. For this reason, it is important 
to define a finite UV quantity in the continuum limit clearly, 
which plays a role of the Zamolodchikov $c$-function in 2-dimensional 
conformal field theory (CFT)~\cite{Zamolodchikov:1986gt}.

Recently, Liu and Mezei proposed such a finite quantity defined from the EE 
in $d$-dimensional quantum field theory (QFT), so-called the renormalized 
entanglement entropy (REE)~\cite{Liu:2012eea}, where 
the size of entangling surface
can be reinterpreted as the renormalization group (RG) flow scale. 
For the conformal fixed points in all dimensions, the REE with 
a spherical entangling surface reproduces the known central charge 
of a given CFT~\cite{Casini:2011kv}. 
Except for the (1+1)-dimensional cases, however,  
understanding the REE along the RG flow still remains to be clarified. 

The REE in (2+1)-dimensional QFT is defined as~\cite{Liu:2012eea}
\begin{align}\label{FSdisk}
{\cal F}_{{\rm disk}}(l) = \left(l \f{\p}{\p l} -1\right)S_{{\rm disk}}(l),
\end{align}
where $S_{{\rm disk}}(l)$ denotes the EE for a disk with radius $l$. 
It was conjectured in \cite{Liu:2012eea} that the REE in 3-dimensions is an analogue of  
$c$-function which counts the number of degrees of freedom at a given energy 
scale, and then ${\cal F}_{{\rm disk}}(l)$ satisfies the so called 
$F$-theorem~\cite{Jafferis:2011zi,Myers:2010tj}. 
This conjecture for any (2+1)-dimensional  
Lorentz invariant field theories was proved in \cite{Casini:2012ei} 
using the strong subadditivity of the EE.   
See also \cite{Klebanov:2012yf,Klebanov:2012va,Ishihara:2012jg} for related works.

There still remains an important issue on the stationarity of the REE at conformal fixed points.
In contrast to the Zamolodchikov $c$-function, 
it was reported that  ${\cal F}_{{\rm disk}}(l)$ of a (2+1)-dimensional free massive field theory {\it would not be stationary}
under a mass deformation \cite{Klebanov:2012va}. 
This result is based on  numerical evaluation of the EE for a disk~\cite{Srednicki:1993im,Huerta:2011qi}. 

Related to the {\it stationarity} issue, 
we take a (2+1)-dimensional interacting CFT,  
the Aharony-Bergman-Jafferis-Maldacena (ABJM) theory~\cite{Aharony:2008ug} including mass deformation \cite{Bena:2004jw,Lin:2004nb}.
In order to investigate the REE, we adopt the holographic method developed in 
\cite{Ryu:2006bv,Ryu:2006ef,Nishioka:2009un}.
It is known that the corresponding dual geometries are given by
the Lin-Lunin-Maldacena (LLM) background with SO(2,1)$\times$SO(4)$\times$SO(4) isometry in 11-dimensional supergravity~\cite{Bena:2004jw,Lin:2004nb},
which have  one-to-one correspondence with the vacua  of the mass-deformed ABJM (mABJM) theory~\cite{Hosomichi:2008jb,Gomis:2008vc}. 
These geometries have been actually conjectured to be 
dual to the supersymmetry 
preserving mass-deformation of the (2+1)-dimensional ${\cal N}=8$ CFT
even before the development of the ABJM theory at the Chern-Simons level $k=1$ \cite{Lin:2004nb}.
See also \cite{SheikhJabbari:2005mf} for the related work. 

In what follows, we review the relation between the vacua of the mABJM theory and 
the LLM geometry, and then compute ${\cal F}_{{\rm disk}}(l)$ near the UV fixed point. 
We analytically compute the small mass effect of the REE for various droplet configurations in the LLM geometry and show 
the {\it stationarity} in (2+1)-dimensional strongly coupled supersymmetric massive
field theory.  Further elaboration for our results
will be presented elsewhere~\cite{Kim:2014qpa}.
 
\section{Vacua of mABJM theory and dual LLM geometry}

The ${\cal N}=6$ mABJM theory has discrete vacua characterized by occupation 
numbers, $N_n$ and $\tilde N_n$, of two types of the Gomis-Rodriguez-Gomez-Van Raamsdonk-Verlinde (GRVV) matrices~\cite{Gomis:2008vc}. 
See also \cite{Kim:2010mr,Cheon:2011gv}.

On the other hand, the LLM geometry under consideration is classified 
by droplet configurations for a given $N$ M2-branes. 
Then $N_n$ and $\tilde N_n$ have one-to-one correspondence 
with the lengths of the black ($l_n$) and white ($\tilde l_n$) droplets in 
the LLM geometry. 
In \cite{Cheon:2011gv},  the explicit mapping  is given to relate the vacua of 
the mABJM theory and the LLM geometry for general $N$ and $k$. 
Therefore, if one picks up one special 
LLM geometry, one can immediately find a corresponding vacuum
in the field theory side.

The LLM geometry with ${\mathbb Z}_k$ quotient is given by
\begin{align}\label{LLMgeom2}
ds^2 &= -e^{4\Phi/3}\left(-dt^2 + dw_1^2 + dw_2^2\right) +e^{-2\Phi/3} h^2(dx^2+dy^2)
\nn \\
&~~~+e^{-2\Phi/3} y e^G ds_{S^3/\mathbb{Z}_k}^2 + e^{-2\Phi/3} y e^{-G} ds_{\tilde S^3/\mathbb{Z}_k}^2
\end{align}
with 
\begin{align}
ds^2_{S^3/\mathbb{Z}_k} &= d\theta^2 + \sin^22\theta \,d\phi^2
+\big((d\lambda + d\varphi/k) + \cos2\theta d\phi\big)^2
\nn \\ \nn
ds^2_{\tilde S^3/\mathbb{Z}_k} &=d\tilde \theta^2 + \sin^22\tilde\theta
\,d\tilde\phi^2
+\big((-d\lambda + d\varphi/k) + \cos2\tilde \theta
d\tilde\phi\big)^2, 
\end{align}
where $e^{-2\Phi} =\mu_0^{-2}(h^2-h^{-2} V^2),\,
h^{-2} = 2 y \cosh G,\, z= \frac12\tanh G$ with a mass parameter $\mu_0$ given
by the transverse 4-form field strength. 
Here, the parameters $k$ and $\mu_0$ 
are identified with the Chern-Simons level and the mass parameter, 
via $\mu_0= \f{\pi m}{2k}$, in the mABJM theory, respectively~\cite{Cheon:2011gv}. 
The LLM geometry \eqref{LLMgeom2} is completely determined in terms of 
$V(x,y)$ and $z(x,y)$, 
\begin{align}
z(x,y) &= \sum_{i=1}^{\infty}\frac{(-1)^{i+1}(x-x_i)}{2\sqrt{(x-x_i)^2 + y^2}},
\nn \\
V(x,y) &= \sum_{i=1}^{\infty}\frac{(-1)^{i+1}}{2\sqrt{(x-x_i)^2 + y^2}},
\end{align}
where $x_i$'s represent the locations of boundary lines between black 
and white strips in the droplet representation. 
See also \cite{Hyun:2013sf} for the large $N$ behavior of the LLM geometry.

\section{Holographic EE for a Disk}
The LLM geometry with ${\mathbb Z}_k$ quotient  
becomes asymptotically AdS$_4\times {\rm S}^7/{\mathbb Z}_k$. 
In the dual field theory side, this implies that the conformal symmetry is restored in
the UV limit.
Therefore, the asymptotic boundary of the LLM geometry is 
the same with that of the original ABJM theory. 
On the other hand, in the IR region, we see the breaking of conformal symmetry 
due to the mass deformation.
This circumstance allows one to investigate the effect
of the mass deformation near 
the UV conformal fixed point of the ABJM theory in the holographic point of view.

The purpose of this paper is to investigate the physical properties of REE 
near the fixed point of LLM geometry.  We note that the value of REE at the UV
fixed point is exactly known by the 
free energy of the ABJM theory~\cite{Casini:2011kv}.
The free energy itself was obtained in terms of the ABJM partition 
function on ${\rm S}^3$ using the localization technique~\cite{Kapustin:2009kz}.
In order to obtain the REE near the UV fixed point of the mABJM theory, 
we first calculate the holographic EE (HEE) for a disk with radius $l$. 
The geometry we are considering 
is the (2+1)-dimensional flat Minkowski space with 7-dimensional compact space 
on the asymptotic boundary. We take a circular region with radius $l$ in the two 
spatial directions of the Minkowski space. 
The expected 9-dimensional surface in HEE proposal is spanned by coordinates, 
$\sigma^i$, $i=1,2,\cdots, 9$. 
Then the induced metric on the surface is given by
\begin{align}\label{indmet}
g_{ij} = \frac{\partial X^M\partial X^N}{\partial \sigma^i\partial\sigma^j} G_{MN},
\end{align}
where $M,N=0,\cdots, 10$ and $G_{MN}$ is the metric of the 11-dimensional  
LLM geometry. 

The proposed HEE~\cite{Ryu:2006bv,Ryu:2006ef} is given by
\begin{align}\label{SAgA}
S_A = \frac{{\rm Min}(\gamma_A)}{4G_N},\quad 
\gamma_A = \int d^9\sigma\, \sqrt{\det g_{ij}},
\end{align}
where $\gamma_A$ is the 9-dimensional surface represented by the induced metric
\eqref{indmet},  ${\rm Min}(\gamma_A)$  the minimum value of $\gamma_A$, 
and $G_N= (2\pi l_{{\rm P}})^9/(32\pi^2)$  the 11-dimensional Newton's constant
with the Planck length $l_{{\rm P}}$.
To compute the HEE for a disk on the asymptotic boundary of the LLM geometry, 
we consider a mapping of coordinates in \eqref{LLMgeom2}, 
\begin{align}\label{mapping1}
&w_1 = \rho \cos \sigma^1, \, w_2 = \rho \sin \sigma^1, r= r(\rho), 
\,\alpha = \sigma^3,
\nn \\
&\theta=\sigma^4,\, \phi=\sigma^5,\,
\tilde\theta=\sigma^6,\, \tilde\phi= \sigma^7,\, \lambda = \sigma^8,~ \varphi = \sigma^9 , 
\end{align}
where $r= \sqrt{x^2 + y^2}$, $\alpha = \tan^{-1}(y/x)$, 
$ 0\le \sigma^1\le 2\pi$, and $ 0\le  \rho(=\sigma^2) \le l$. 
Applying the mapping \eqref{mapping1} to \eqref{SAgA} and integrating out 
the coordinates of the two $S^3$'s and $\sigma^1$, we obtain 
\begin{align}\label{gammaA}
\gamma_A
=\frac{\pi^5 R^9}{16 k \mu_0}\int_0^l d\rho \,\int_0^{\pi} d\alpha \, 
\frac{f \rho  \sin ^2 \alpha}{u^3}
\sqrt{1+ \frac{f^2  u'^2}{4 \mu_0^2 \sin^2 \alpha   u^2}},
\end{align}
where $u = \frac{R^3}{4r}$ with 
the radius of the ${\rm AdS}_4$, $R= (32\pi^2 k N)^{1/6} l_{\rm P}$, and 
\begin{align}\label{fralp}
f(u,\alpha) = \sqrt{1- 4\tilde z^2 - 4\tilde y^2 \tilde V^2}. 
\end{align}
Here we rescaled coordinates and functions as 
\begin{align} \tilde x = \frac{4 x}{R^2},\,\,\, 
\tilde y = \frac{4 y}{R^2}, \,\,\, \tilde V(\tilde x,\tilde y) =\frac{R^2}{4} V(x,y) .
\end{align}
Under these rescalings, $z(x,y)$ does not change. 

As shown  in \eqref{gammaA}, the surface area $\gamma_A$ depends on the function 
$f(u, \alpha)$ which includes all information about possible droplets of the LLM 
geometry. 
Therefore, in order to figure out the properties of the HEE for a given droplet configuration, 
one can analyze $f(u,\alpha)$ in various limits. 
In this paper, we are interested in the physics near the UV fixed point, 
and so it is enough to focus on the properties of $\gamma_A$ 
for a small mass deformation. 

The gauge/gravity duality implies that there is a correspondence between a strongly 
coupled gauge theory and a weakly curved gravity. 
As is well known,
for the validity of this correspondence, one needs to take the large $N$ limit.
In our case, however, this large $N$ limit 
is not enough to have well defined HEE:
For a given $N$,  the possible number of distinguishable LLM geometries is given by
the partition of $N$, $p(N)$, and it behaves as 
$p(N)\sim e^{\pi\sqrt{\frac{2N}{3}}}$ in the large $N$ limit. 
It is not sure that the geometries for all such possibilities are weakly curved over 
the whole space-time region.
Thus it is necessary to take a step of selecting suitable geometries which are weakly
curved in the large $N$ limit. 
Fortunately, there is a guideline to select weakly curved geometries.
In the Young diagram representation of the LLM geometry, the lengths of
the white/black strips are mapped to those of the 
horizontal/vertical edges of the Young diagram.  The area of the diagram is identified with $N$. 
For geometries without strongly curved region, one has to consider the Young
diagram including only the long edges which are of the order of 
$\sqrt{N}$~\cite{Lin:2004nb}.
If some of the edges are short ($\ll \sqrt{N}$), then their
presence makes the geometry highly curved.
This characteristic feature of the LLM geometries has been observed in \cite{Hyun:2013sf}. 
From the curvature behavior of the LLM geometry, we conclude that the geometry 
corresponding to the rectangular shaped Young diagram with the side length of order $\sqrt{N}$  is weakly curved over the whole space-time region~\cite{Hyun:2013sf}. In this case 
the gauge/gravity duality is well-defined.
From now on, we will concentrate on the LLM geometry
having such rectangular shaped Young diagram. 

Under the small mass deformation, we expand $f(u, \alpha)$ in \eqref{fralp} 
for a rectangular shaped Young diagram with side lengths, 
denoted by $w$ and $b$,  as  
\begin{align}
&f(u , \alpha) =  2\sin\alpha (\mu_0 u)\Big[1+\frac{\tilde\sigma\cos\alpha}{\sqrt{2}} 
(\mu_0 u)
\nn \\
&+\frac{\tilde\sigma^2-1
+(5\tilde\sigma^2 + 9)\cos 2\alpha}{8} (\mu_0 u)^2 
+{\cal O}\left((\mu_0 u)^3\right)  \Big],
\end{align}
where $w = \frac{\sqrt{kN}}{\hat\sigma}$, $b = \hat\sigma\sqrt{kN}$, 
and $\tilde\sigma = \hat\sigma - \frac1{\hat\sigma}$. 
For the fully symmetric case, $\hat\sigma = 1$ and hence $\tilde\sigma=0$.  
As discussed in the previous paragraph, $f(u, \alpha)$ is valid for
$|\tilde{\sigma}| \ll \sqrt{kN}$.

After the integration over $\alpha$,  
the surface area $\gamma_A$ is expanded in terms of  small mass as follows: 
\begin{align}\label{act:deformed}
&\gamma_A = \frac{\pi^5 R^9}{6 k} \int_0^l d\rho  \,
\rho\, \bigg[\frac{1}{u^2}  \sqrt{1 + u'^2} - \frac{5\tilde\sigma^2 + 16}{20\left( 1 + u'^2  \right)^{3/2}}\,\, \mu_0^2
\nn \\
&- \frac{ 3(\tilde\sigma^2+4)(3 u'^2 + 2 u'^4)}{
5\left( 1 + u'^2  \right)^{3/2}}\,\, \mu_0^2 +\cdots \bigg] .
\end{align}
The minimum value of $\gamma_A$ 
is given by the solution of the equation of motion 
for $u(\rho)$ after regarding $\gamma_A$ as a classical Euclidean action. 
In the $\mu_0\to 0$ limit, $\gamma_A$ is reduced to the case of the AdS$_4$ up to 
an overall factor coming from the contribution of the 7-dimensional compact space. 
In this limit, a special solution is known,
satisfying the appropriate boundary conditions, $u_0'(0)=0$ and $u_0(l)=0$~\cite{Ryu:2006ef},
\begin{align}
u_0(\rho) = \sqrt{l^2 - \rho^2} .
\end{align}
In order to see the effect of mass deformation near the UV fixed point, 
we consider the perturbation with the mass parameter $\mu_0$ around the 
solution $u_0(\rho)$. 
Since the first correction appears at $\mu_0^2$ order and we are 
interested only in the leading order correction to $\gamma_A$, we take 
\begin{align}\label{pertu}
u(\rho) = u_0(\rho)+ (\mu_0 l)^2  \delta u(\rho)  .
\end{align} 
From the equation of motion for $u(\rho)$, it is possible to get a general solution 
of $\delta u(\rho)$.
By imposing two boundary conditions, $\delta u'(0)=0$ and $\delta u(l)=0$, 
two integration constants are fixed and what we obtain at the end is 
\begin{align}\label{delta}
\delta u(\rho)  &= \frac{l^3}{300\sqrt{1- (\rho/l)^2}}\Big[206 \tilde\sigma^2 + 800\nn \\
&\times \Big(\tanh^{-1}\sqrt{1- (\rho/l)^2}  + \ln (\rho/l)
-\sqrt{1- (\rho/l)^2} \Big) 
\nn \\
&+\frac{81\tilde\sigma^2}{2} \left(\rho/l\right)^6 
+\frac{160 -11 \tilde\sigma^2}{2} \left(\rho/l\right)^4 
\nn \\
&-(560 + 119\tilde\sigma^2)(\rho/l)^2 
+\frac{1120 + 281\tilde\sigma^2}{2}\Big].
\end{align}
\begin{figure}
\begin{center}
\includegraphics[width=6cm,clip]{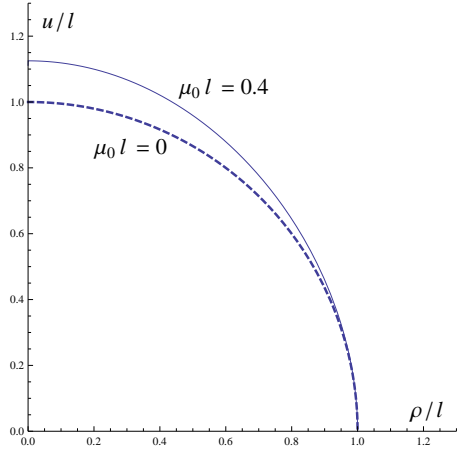}
\end{center}
\caption{
\label{fig:msurface}
Change of the minimal surface due to the mass deformation
in the symmetric case $\tilde\sigma=0$}
\end{figure}
The effect of mass deformation on the minimal surface is depicted in 
Fig.~\ref{fig:msurface}. 

Plugging \eqref{pertu} into \eqref{act:deformed} and performing the 
$\rho$ integration, 
we obtain the HEE from \eqref{SAgA}
for a disk with radius $l$, up to $\mu_0^2$ order, 
\begin{align}\label{Sdisk1}
S_{{\rm disk}} = \frac{\pi ^5  R^9}{12 k G_N}\left(\frac{l}{\epsilon} -1 
-\frac{103\tilde\sigma^2 + 400}{300}\,(l \mu_0)^2\right),
\end{align}
where the constant $\epsilon$ is the UV cutoff in the $u$ coordinate. 
The first two terms correspond to the HEE of the
ABJM theory without mass deformation.   The last term which is one of our main 
results is the leading correction from the small mass expansion. 
If the idea of HEE is correct, 
the above expression is interpreted as the EE for the strongly interacting  
massive field theory.  Although it is hard to compute the EE on the field theory side, 
it would be desirable to check whether the HEE of (\ref{Sdisk1})
matches with the would-be field theory result.

\section{REE and $c$-function in (2+1)-dimensions}

From the HEE in \eqref{Sdisk1} for the rectangular droplets, one obtains
the REE defined in \eqref{FSdisk}, up to $\mu_0^2$ order,
\begin{align}\label{Fdisk3}
{\cal F}_{{\rm disk}}(l) =F- 
\frac{\pi ^5  R^9(103\tilde\sigma^2 + 400 )}{3600 kG_N}\,(l \mu_0)^2,
\end{align}
where $F = \frac{\pi ^5  R^9}{12 k G_N}$ is the free energy of the 
ABJM theory without mass deformation. As we explained previously, the REE can 
be a $c$-function in
holographic point of view, 
which counts the number of effective degrees of freedom of a given system 
at the length scale $l$. 
The REE in \eqref{Fdisk3} shows that the expected holographic 
$c$-function near the UV fixed point is positive and monotonically 
decreases along the RG flow as the system size $l$ increases. 
This result supports the $F$-theorem in 3-dimensional field 
theory~\cite{Jafferis:2011zi}. 
The REE in \eqref{Fdisk3} describes the behavior of the $c$-function around 
the UV fixed point. 

The conformal symmetry of the ABJM theory is broken by perturbing in terms of 
the fermonic mass term~\cite{Hosomichi:2008jb,Gomis:2008vc,Bena:2000zb,Lambert:2009qw}, 
\begin{align}
{\cal L}_{{\rm ferm}}\sim \mu_0 \Psi^{\dagger A} M_A^B \Psi_B
\end{align} 
with the mass matrix $M_A^B = {\rm diag}(1,1,-1,-1)$.  Then the dimensionless 
coupling constant for the relevant mass deformation is $g= l\mu_0$, and 
the corresponding coupling constant for the bosonic scalar fields is proportional to $g^2$ 
due to the supersymmetry.  
As for the stationarity issue at the UV fixed point in (2+1)-dimensional interacting 
field theory~\cite{Klebanov:2012va}, 
we note that  the REE of (\ref{Fdisk3}) leads to\footnote{This result is the check for a supersymmetric mass deformation using analytic calculations, 
but it is not clear whether non-supersymmetric mass 
deformations~\cite{Klebanov:2012va} guarantees or not the stationarity in strongly coupled models. 
We need more investigations in this direction.}
\begin{align}
\frac{\partial {\cal F}_{{\rm disk}}}{\partial g}\Big|_{g=0} = 0.
\end{align}
This result shows that the REE in (2+1)-dimensional CFT 
is  stationary under the relevant mass deformation, 
like the $c$-function in (1+1)-dimensions~\cite{Zamolodchikov:1986gt} 
and the $a$-function in (3+1)-dimensions~\cite{Cardy:1988cwa}.

\section{Summary}

In this paper we have investigated the UV behavior of the REE for a disk 
in (2+1)-dimensional field theory. 
In holographic point of view we have confirmed analytically that 
the REE is positive,  monotonically decreasing along the RG flow, and 
stationary under a relevant mass deformation at the UV conformal fixed point. 
As a top-down approach for the gauge/gravity duality, we have contemplated on the 11-dimensional 
LLM geometries dual to the vacua of the mABJM theory. 
Our analytic results can be applied to various droplet configurations 
in the LLM geometry. 

Near the UV fixed point, 
the roles of the REE are the same with the Zamolodchikov $c$-function in 2-dimensions and 
the $a$-function in 4-dimensions.  
We would like to emphasize that we have considered an interacting theory 
and given an analytic result. Our computations show that the REE, 
which counts the effective degrees of freedom at a given length (energy) scale,  
can be a strong candidate for the $c$-function in 3-dimensions.

As a final remark, the investigation of the IR region for the mABJM theory will certainly give some insights on other aspects of the REE.

\section*{Acknowledgments} 
We would like to thank M. Mezei for helpful comments on the stationarity. 
This work was supported by the Korea Research
Foundation Grant funded by the Korean Government
with Grant No. 2011-0009972 (O.K.),  NRF-2013R1A1A2A10057490 (C.P.), 
and NRF-2012R1A1A2004203, 2012-009117, 2012-046278 (H.S.). 
It was also supported by the World Class University Grant No. R32-10130 (O.K. and C.P.) 
and the Basic Science Research Program through the National Research Foundation of Korea(NRF) funded by the Ministry of Science, ICT $\&$ Future Planning(2014R1A1A1003220) (K.K.).



\begin{thebibliography}{99}


\bibitem{Bombelli:1986rw} 
  L.~Bombelli, R.~K.~Koul, J.~Lee and R.~D.~Sorkin,
  ``A Quantum Source of Entropy for Black Holes,''
  Phys.\ Rev.\ D {\bf 34}, 373 (1986).
  
\bibitem{Srednicki:1993im} 
  M.~Srednicki,
  ``Entropy and area,''
  Phys.\ Rev.\ Lett.\  {\bf 71}, 666 (1993)
  [hep-th/9303048].
  
\bibitem{Zamolodchikov:1986gt} 
  A.~B.~Zamolodchikov,
  ``Irreversibility of the Flux of the Renormalization Group in a 2D Field Theory,''
  JETP Lett.\  {\bf 43}, 730 (1986)
  [Pisma Zh.\ Eksp.\ Teor.\ Fiz.\  {\bf 43}, 565 (1986)].
  
 
\bibitem{Liu:2012eea} 
  H.~Liu and M.~Mezei,
  ``A Refinement of entanglement entropy and the number of degrees of freedom,''
  JHEP {\bf 1304}, 162 (2013)
  [arXiv:1202.2070 [hep-th]].
  
  
\bibitem{Casini:2011kv} 
  H.~Casini, M.~Huerta and R.~C.~Myers,
  ``Towards a derivation of holographic entanglement entropy,''
  JHEP {\bf 1105}, 036 (2011)
  [arXiv:1102.0440 [hep-th]].

\bibitem{Jafferis:2011zi} 
  D.~L.~Jafferis, I.~R.~Klebanov, S.~S.~Pufu and B.~R.~Safdi,
  JHEP {\bf 1106}, 102 (2011)
  [arXiv:1103.1181 [hep-th]].

  
\bibitem{Myers:2010tj} 
  R.~C.~Myers and A.~Sinha,
  Phys.\ Rev.\ D {\bf 82}, 046006 (2010)
  [arXiv:1006.1263 [hep-th]];
  JHEP {\bf 1101}, 125 (2011)
  [arXiv:1011.5819 [hep-th]].
   
   
\bibitem{Casini:2012ei} 
  H.~Casini and M.~Huerta,
  Phys.\ Rev.\ D {\bf 85}, 125016 (2012)
  [arXiv:1202.5650 [hep-th]].
  
 
\bibitem{Klebanov:2012yf} 
  I.~R.~Klebanov, T.~Nishioka, S.~S.~Pufu and B.~R.~Safdi,
  JHEP {\bf 1207}, 001 (2012)
  [arXiv:1204.4160 [hep-th]].
  
\bibitem{Klebanov:2012va} 
  I.~R.~Klebanov, T.~Nishioka, S.~S.~Pufu and B.~R.~Safdi,
  JHEP {\bf 1210}, 058 (2012)
  [arXiv:1207.3360 [hep-th]].
  
  
\bibitem{Ishihara:2012jg} 
  M.~Ishihara, F.~-L.~Lin and B.~Ning,
  Nucl.\ Phys.\ B {\bf 872}, 392 (2013)
  [arXiv:1203.6153 [hep-th]];
  T.~Nishioka and K.~Yonekura,
  JHEP {\bf 1305}, 165 (2013)
  [arXiv:1303.1522 [hep-th]];
  Y.~Bea, E.~Conde, N.~Jokela and A.~V.~Ramallo,
  JHEP {\bf 1312}, 033 (2013)
  [arXiv:1309.4453 [hep-th]];
  H.~Liu and M.~Mezei,
  JHEP {\bf 1401}, 098 (2014)
  [arXiv:1309.6935 [hep-th].
  
\bibitem{Huerta:2011qi} 
  M.~Huerta,
  Phys.\ Lett.\ B {\bf 710}, 691 (2012)
  [arXiv:1112.1277 [hep-th]];
  H.~Casini and M.~Huerta,
  J.\ Phys.\ A {\bf 42}, 504007 (2009)
  [arXiv:0905.2562 [hep-th]];
I. Peschel, 
Journal of Physics A Mathematical General 36,  L205, arXiv:cond-mat/0212631.

\bibitem{Aharony:2008ug} 
  O.~Aharony, O.~Bergman, D.~L.~Jafferis and J.~Maldacena,
  JHEP {\bf 0810}, 091 (2008)
  [arXiv:0806.1218 [hep-th]].


\bibitem{Bena:2004jw} 
  I.~Bena and N.~P.~Warner,
  JHEP {\bf 0412}, 021 (2004)
  [hep-th/0406145].

\bibitem{Lin:2004nb} 
  H.~Lin, O.~Lunin and J.~M.~Maldacena,
  JHEP {\bf 0410}, 025 (2004)
  [hep-th/0409174].

\bibitem{Ryu:2006bv} 
  S.~Ryu and T.~Takayanagi,
  Phys.\ Rev.\ Lett.\  {\bf 96}, 181602 (2006)
  [hep-th/0603001].
  
\bibitem{Ryu:2006ef} 
  S.~Ryu and T.~Takayanagi,
  JHEP {\bf 0608}, 045 (2006)
  [hep-th/0605073].
  
\bibitem{Nishioka:2009un} 
  T.~Nishioka, S.~Ryu and T.~Takayanagi,
  J.\ Phys.\ A {\bf 42}, 504008 (2009)
  [arXiv:0905.0932 [hep-th]];
  
  T.~Takayanagi,
  Class.\ Quant.\ Grav.\  {\bf 29}, 153001 (2012)
  [arXiv:1204.2450 [gr-qc]].

\bibitem{Hosomichi:2008jb} 
  K.~Hosomichi, K.~-M.~Lee, S.~Lee, S.~Lee and J.~Park,
  JHEP {\bf 0809}, 002 (2008)
  [arXiv:0806.4977 [hep-th]].


\bibitem{Gomis:2008vc} 
  J.~Gomis, D.~Rodriguez-Gomez, M.~Van Raamsdonk and H.~Verlinde,
  JHEP {\bf 0809}, 113 (2008)
  [arXiv:0807.1074 [hep-th]].
 
\bibitem{SheikhJabbari:2005mf} 
  M.~M.~Sheikh-Jabbari and M.~Torabian,
  JHEP {\bf 0504}, 001 (2005)
  [hep-th/0501001].
 
 
\bibitem{Kim:2014qpa} 
  K.~K.~Kim, O.~K.~Kwon, C.~Park and H.~Shin,
  arXiv:1407.6511 [hep-th].
  
    
\bibitem{Kim:2010mr} 
  H.~-C.~Kim and S.~Kim,
  Nucl.\ Phys.\ B {\bf 839}, 96 (2010)
  [arXiv:1001.3153 [hep-th]].
   

  

  
\bibitem{Cheon:2011gv} 
  S.~Cheon, H.~-C.~Kim and S.~Kim,
  arXiv:1101.1101 [hep-th].
  
  
\bibitem{Hyun:2013sf} 
  Y.~-H.~Hyun, Y.~Kim, O-K.~Kwon and D.~D.~Tolla,
  Phys.\ Rev.\ D {\bf 87}, no. 8, 085011 (2013)
  [arXiv:1301.0518 [hep-th]].
  
\bibitem{Kapustin:2009kz} 
  A.~Kapustin, B.~Willett and I.~Yaakov,
  JHEP {\bf 1003}, 089 (2010)
  [arXiv:0909.4559 [hep-th]].
  
  
  
\bibitem{Bena:2000zb} 
  I.~Bena,
  Phys.\ Rev.\ D {\bf 62}, 126006 (2000)
  [hep-th/0004142].
  
\bibitem{Lambert:2009qw} 
  N.~Lambert and P.~Richmond,
  JHEP {\bf 0910}, 084 (2009)
  [arXiv:0908.2896 [hep-th]].
  
\bibitem{Cardy:1988cwa} 
  J.~L.~Cardy,
  Phys.\ Lett.\ B {\bf 215}, 749 (1988);
  Z.~Komargodski and A.~Schwimmer,
  JHEP {\bf 1112}, 099 (2011)
  [arXiv:1107.3987 [hep-th]];
  Z.~Komargodski,
  JHEP {\bf 1207}, 069 (2012)
  [arXiv:1112.4538 [hep-th]];
  M.~A.~Luty, J.~Polchinski and R.~Rattazzi,
  JHEP {\bf 1301}, 152 (2013)
  [arXiv:1204.5221 [hep-th]].
  
   

  
\end{thebibliography}
\end{document}